\documentclass[prl,reprint, nofootinbib,amsmath,amssymb, aps, floatfix]{revtex4-1}
\usepackage{dcolumn}
\usepackage{bm}
\usepackage{amssymb,amsmath,amsthm,graphicx,ulem}
\usepackage[linktoc=page]{hyperref}

\usepackage{graphicx,subfigure}
\usepackage{epsfig}
\usepackage{amsmath}
\usepackage{amsfonts}
\usepackage{amssymb}
\usepackage[usenames]{color}
\usepackage{enumerate}
\usepackage{bold-extra}

\usepackage[letterpaper,left=2.cm,right=2.cm,top=2.5cm,bottom=2.5cm]{geometry}
\usepackage[T1]{fontenc}

\DeclareSymbolFont{matha}{OML}{txmi}{m}{it}
\DeclareMathSymbol{v}{\mathord}{matha}{118}

\begin{document}
\normalem

\title{Decoding the Apparent Horizon:\\ A Coarse-Grained Holographic Entropy}
\author{Netta Engelhardt}
\affiliation{Department of Physics, Princeton University, Princeton NJ 08544 USA}
\email{nengelhardt@princeton.edu}
\author{Aron C. Wall}%
\affiliation{Institute for Advanced Study, Einstein Drive, Princeton NJ 08540 USA}
\email{aroncwall@gmail.com}

\bibliographystyle{utcaps}

\begin{abstract}
When a black hole forms from collapse in a holographic theory, the information in the black hole interior remains encoded in the boundary. We prove that the area of the black hole's apparent horizon is precisely the entropy associated to coarse graining over the information in its interior, subject to knowing the exterior geometry. This is the maximum holographic entanglement entropy that is compatible with all classical measurements conducted outside of the apparent horizon. We identify the boundary dual to this entropy and explain why it obeys a Second Law of Thermodynamics.
\end{abstract}

\maketitle

\section{Introduction}

The Second Law of Thermodynamics states that entropy increases with time. One natural notion of entropy is the von Neumann entropy:
\begin{equation}
S[\rho]=-\mathrm{tr}(\rho \ln \rho),
\end{equation}
where $\rho$ is the density matrix of a quantum system. However, this quantity is conserved under unitary time evolution, in apparent tension with the Second Law. To obtain an increasing entropy, it is necessary to
coarse grain $S$ by ``forgetting'' certain information, since the vast majority of microscopic data in a thermal system is inaccessible to macroscopic observations. One common coarse-graining method is the maximization of the system's entropy subject to fixed the values of a set of feasible macroscopic measurements $\mathcal{M}(t)$ at a moment in time \cite{Jay57a, Jay57b,GelHar06}:

\begin{equation}
S^{\mathrm{coarse}}(t) = \max\limits_{\rho'}\left (S[\rho'] : \mathcal{M}(t) \right).
\end{equation}
Assuming that any ordered information inaccessible at early times remains so at later times, $S^{\mathrm{coarse}}$ should increase with time, defining a nontrivial Second Law.

The most mysterious application of the Second Law is to black holes. Stationary black holes (e.g. Kerr) have entropy, which is proportional to the area of their horizon $H$~\cite{Bek72, Haw75}:
\begin{equation}
S_{BH} = \frac{\mathrm{Area}[H]}{4G\hbar},
\end{equation}
as suggested by the Laws of Black Hole Mechanics~\cite{Haw71, BarCar73, Bek72, Haw75}. However, despite some clues from string theory and other approaches
(reviewed in \cite{Car08}),
it is still unclear in general what microscopic quantum-gravitational degrees of freedom are counted by this entropy. Dynamically evolving black holes such as those formed from stellar collapse are even more controversial, since there are multiple possible definitions of a horizon, e.g. the event horizon and the apparent horizon~\cite{HawEll} --- and correspondingly, multiple area increase theorems~\cite{Haw71, Hay93, AshKri02, GouJar06, BouEng15a, SanWei16}.

In holographic models of quantum gravity, a black hole is dual to some boundary state $\rho$ whose von Neumann entropy $S[\rho]$ can be computed from a compact extremal (HRT) surface in the bulk, as conjectured in~\cite{RyuTak06, HubRan07} and essentially proven in~\cite{LewMal13,DonLew16}:
\begin{equation}
S[\rho]= \frac{\mathrm{Area}[X_{HRT}]}{4G\hbar}.
\end{equation}
A surface is extremal if its area is unchanged by any first order perturbation to the surface's location; if there is more than one, $X_{HRT}$ is the one with the minimal area extremal surface (and homologous to the boundary~\cite{HubRan07, HeaTak}). \footnote{Because we restrict attention to the entropy of the whole CFT, in this case the HRT surfaces are compact and do not reach the boundary.} This quantity is independent of time, so it is not suitable for describing the entropy increase of a growing black hole. 
Unitarity of the boundary theory implies that no information is lost, but this is not enough: to account for the increase of black hole entropy, a coarse graining scheme must be specified.

Even though black hole thermodynamics was the original motivation for the holographic principle~\cite{Tho93, Sus95}, no one has yet given a clear explanation of the role of the black hole horizon as a repository of information about the interior.
Indeed, it was recently shown~\cite{EngWal17} that if we know the outcome of all classical measurements $\mathcal{M}(t)$ outside of the event horizon $H$, then $S^{\mathrm{coarse}}<\mathrm{Area}[H]/4G\hbar$: we have access to too much information for our remaining ignorance to be given by the event horizon's area (thus refuting a broad class of proposals relating entropy to area, including \cite{Sor97, BiaMye12, FreMos13, KelWal13}.)

We therefore look for alternatives to the event horizon. An appealing option is the apparent horizon $\mu$, the outermost compact surface (at a moment of time) which is marginally outer trapped \cite{HawEll}, i.e. the the expansion $\theta_k \equiv \nabla_k \ln(\mathrm{Area}[n]) = 0$, where $k$ is a future-outwards null vector, and $n$ is a small pencil of lightrays shot out in the $k$-direction from a small neigborhood of a point on $\mu$.
In the case of a black hole that forms from collapse, such marginally trapped surfaces form behind the event horizon, even though the HRT surface is the empty set (so that the boundary state is pure).

In this Letter, we give a geometric proof (using classical GR methods in the bulk) that the area of the apparent horizon $\mu$ \textit{does} play the role of a coarse-grained entropy:
\begin{equation}
S^{\mathrm{coarse}} = \frac{\mathrm{Area}[\mu]}{4G\hbar},
\end{equation}
where we coarse grain over the region behind the apparent horizon (the ``microstates'') while holding all classical measurements in the exterior fixed (i.e. we fix all data in the exterior, but working in the classical regime).  This makes it plausible that the interior is encoded holographically by a set of independent qubits, one per $4 / \ln 2$ Planck-areas, on the apparent horizon (but not the event horizon!) \cite{Bou99d, Bou02, SanWei16a, NomSal16b}.  Our classical proof explicitly constructs the entropy-maximizing geometry, which would correspond to maximally scrambling all of these qubits.  If our result can be extended to the quantum regime (along the lines of \cite{Wal10QST, BouEng15c,EngWal14,FauLew13,DonLew17}
it might provide insight into the firewalls paradox \cite{AMPS, AMPSS,MatPlu11,BraPir09}, a puzzle about whether maximally scrambled black holes have an interior. An investigation on areas of non-compact analogues of the apparent horizons will appear in~\cite{MarWhiTA}.

Note that although apparent horizons are highly non-unique due to the choice of time slicing, the above construction is valid for each of them. 

We also identify the boundary dual to $S^{\mathrm{coarse}}$ of the apparent horizon. This quantity may be computed by maximizing the boundary von Neumann entropy while keeping fixed the outcomes of a set of ``simple'' experiments performed \textit{after} a given moment in time. This new entry in the holographic dictionary (which we show is exact to all orders in perturbation theory for near-equilibrium black holes), extends the HRT prescription to a much more general class of bulk surfaces.

Both the bulk and corresponding boundary entropies automatically satisfy the Second Law.  This provides the first valid holographic explanation of the Area Increase Law for black holes.

\section{Outer Entropy}
The outer entropy is a coarse-grained entropy that holds fixed the exterior of a codimension-2 surface $\sigma$.  We define $O_{W}[\sigma]$, the outer wedge, as the region spacelike outside of $\sigma$ (on the side with the asymptotic boundary). The outer entropy is
\begin{equation}
S^{\mathrm{(outer)}}[\sigma] \equiv \max\limits_{\rho'}\left ( S[\rho']: O_{W}[\sigma] \right),
\end{equation}
where $\rho'$ is any state of the boundary CFT with a classical bulk dual geometry $M'$; we choose $\rho'$ to maximize the von Neumann entropy $S[\rho'] = X_{HRT}[M']$, subject to the constraint that $M'$ have the same outer wedge $O_{W}[\sigma]$ as the original classical bulk $M$ dual to $\rho$.  Although we have phrased this maximization in terms of the boundary state, note that this can be regarded as a pure bulk construction involving maximizing the area of the HRT surface. The only holographic aspect (in this section) is the identification of an extremal surface lodged inside the black hole with a fine-grained entropy (i.e. the von Neumann entropy). Any theory with such an identification --- even one with asymptotically flat boundary conditions (should such a theory exist) --- allows the interpretation of $S^{\mathrm{(outer)}}$ as a coarse-grained entropy.

While this coarse-grained entropy can be defined for a general surface $\sigma$, when $\sigma=\mu$, an apparent horizon, we will show that:
\begin{equation}\label{eq:SisA}
S^{\mathrm{(outer)}}[\mu]=\frac{\mathrm{Area}[\mu]}{4 G \hbar}.
\end{equation}
Hence, the area of the apparent horizon has a statistical interpretation as \emph{the maximum boundary entropy that is compatible with the geometry of its exterior}.  This provides a holographic answer to the disputed question: what does the Bekenstein-Hawking entropy of a black hole count? \cite{Jac99, Sor05, JacMar05, FreHub05, HsuRee08, Mar08} \\

\textbf{Outline of Proof:}
Let $k$ (respectively $\ell$) be the orthogonal future-directed null vectors pointing outward (respectively inward) from a surface.  An extremal surface $X$ satisfies $\theta_k = \theta_\ell = 0$.  An HRT surface additionally must be the minimal area surface (homologous to the boundary) on some spatial slice $\Sigma$ \cite{Wal12}.

An apparent horizon $\mu$ (an outermost marginally trapped surface) satisfies $\theta_k = 0$, $\theta_\ell \le 0$, and (generically) $\nabla_k \theta_\ell < 0$ \cite{AndMet07,Mar14}.  We assume that $\mu$ is homologous to the boundary, i.e. there exists a spatial slice $\Sigma$ connecting $\mu$ to the boundary, and moreover that there exists a $\Sigma$ such that the area of any surface circumscribing $\mu$ is larger than the area of $\mu$. These requirements are reasonable for black hole horizons.

In any spacetime, $\mathrm{Area}[X_{HRT}]\le \mathrm{Area}[\mu]$; this can be proven by a simple focusing argument: in a spacetime satisfying the Null Energy Condition ($T_{vv} \ge 0$ for any null vector $v$), a null surface $N_{\pm k}[\mu]$ shot out along the $\pm k$-direction of $\mu$ has monotonically decreasing area moving away from $\mu$ along $N_{\pm k}$ in the $+k$ or $-k$ directions, where we truncate the surface when generators intersect \cite{Pen65,HawEll,Wald}.  We extend $N_{\pm k}$ along its generators to the slice $\Sigma$ on which $X_{HRT}$ is minimal \cite{Wal12}.
\begin{equation}
\mathrm{Area}[\mu] \ge \mathrm{Area}[\Sigma\cap N_{\pm k}[\mu]] \ge \mathrm{Area}[X_{HRT}].
\end{equation}
Hence the entropy $S[\rho']$ cannot exceed $\mathrm{Area}[\mu]/4G\hbar$.

To prove that this inequality is saturated, we construct a bulk spacetime $M'$ (with the same outer wedge $O_{W}[\mu]$) satisfying Area$[X_{HRT}]=\mathrm{Area}[\mu]$.  To specify the interior data in $M'$, we impose initial data on $N_{-k}$, the null surface fired from $\mu$ in the $-k^{a}$ direction.  We choose our initial data so that the surface $N_{-k}$ is \emph{stationary}; every cross-section has the same geometry.  (The Appendix shows this construction satisfies the constraint equations, so that a spacetime solution $M'$ exists, due to $\nabla_{\ell}\theta_{k}<0$.)

\begin{figure}[ht]
\includegraphics[width=0.35 \textwidth]{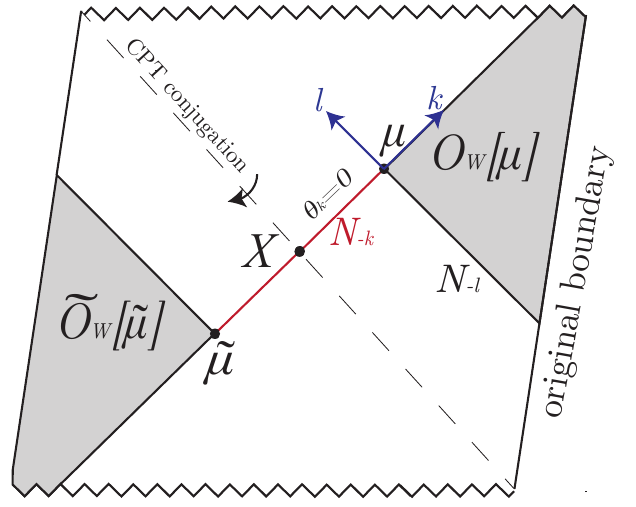}
\caption{The coarse-grained spacetime dual to the state $\rho'$ with maximal $S[\rho']$ and fixed $O_{W}[\mu]$ (shaded gray). The null congruence $N_{-k}$ (red) is fired from $\mu$ towards the $-k$ direction and is stationary. The congruence $N_{-l}$, the past boundary of $O_{W}[\mu]$, is fired in the $-\ell$ direction from $\mu$. $X$ is the HRT surface of the coarse-grained spacetime. Tilded quantities represent the CPT mirror reverse.}
\label{fig:construction}
\end{figure}

By following $N_{-k}$ far enough, we eventually come to an extremal surface $X$ (see Appendix for details).  Since $N_{-k}$ is stationary, Area$[X]=\mathrm{Area}[\mu]$.  We can complete the spacetime by requiring it to be invariant under a CPT-reflection about $X$ (i.e. we reflect space and time about $X$ while exchanging matter with antimatter).  See Fig.~\ref{fig:construction}.  The resulting bulk $M'$ has two asymptotic boundaries, and therefore represents a pure state (analogous to the thermofield double wormhole construction~\cite{Mal01}).  When the state $\rho'$ is restricted to a single boundary, the entropy $S[\rho']=X_{HRT}[M']$.  (Note that the region $O_{W}[X_{HRT}]$ 
agrees with the original bulk geometry dual to $\rho$~\cite{CzeKar12, Wal12, HeaHub14, JafLew15, DonHar16,FauLew17}.)

Because $N_{-k}$ is stationary and by assumption $\mu$ is minimal on a slice of $O_{W}[\mu]$, we now have an initial data slice $\Sigma$ on which $X$ is the minimal cross-section.  Any other extremal surface $X'$ has greater area than $X$:
\begin{equation}
\mathrm{Area}[X'] \ge \mathrm{Area}[\Sigma\cap N_{\pm k}[X']] \ge \mathrm{Area}[X],
\end{equation}
where the first inequality comes from focusing of a null surface $N_{\pm k}[X']$ shot out from $X'$.
Hence $X = X_{HRT}$, proving Eq.~\eqref{eq:SisA}.

\section{Simple Entropy}
Thus far, our coarse-grained entropy has been defined from the bulk point of view. We now identify the boundary dual to the outer entropy, which we call the \textit{simple entropy}, as it relies on ``simple operators''.

In AdS/CFT, single trace operators on the boundary correspond to locally propagating fields in the bulk. More generally, we expect that the product of a small number of single trace operators also propagates locally in the bulk. However, it is known that sufficiently complicated operators (known as precursors~\cite{PolSus99, Fre02}) can change the deep bulk region acausally; hence to define a coarse graining that is dual to $O_{W}[\mu]$, we must avoid such complicated operations. We therefore define a ``simple'' experiment as a procedure performed after a moment of time $t_{i}$, in which we measure a local operator $\mathcal{O}(t>t_{i})$ after having turned on a set of local sources $J(t>t_{i})$; we require that these sources propagate causally into the bulk. For classical solutions, we can restrict attention to one-point operators and sources, since the higher-point functions are determined from them.  (The ``one-point entropy'' \cite{KelWal13}, proposed as a holographic dual to the area of the event horizon, did not allow sources.)  To prevent recurrences, we implicitly include a late time cutoff $t_{f}$ prior to exponentially large values of $t$.


The simple entropy is now defined as the maximum entropy of a state $\rho'$ compatible with the outcomes of all such simple experiments (i.e. the maximization is done over a subspace of $\rho$'s that all yield the same outcomes):
\begin{equation}\label{eq:simple}
S^{\mathrm{(simple)}}(t_{i})= \max\limits_{\rho'}\left (  S[\rho'] : \langle E^{\dagger} \mathcal{O}(t) E\rangle \ \mathrm{fixed} \right),
\end{equation}
where $\rho'$ is defined at $t_{i}$, and
\begin{equation} E=\mathcal{T}\mathrm{exp}[-i \int\limits_{t_{i}}^{t}J(t')\ \mathcal{O}_{J}(t')\ dt']
\end{equation}
is the time-ordered insertion of sources $J(t)$ used to prepare the simple experiment by which $\mathcal{O}(t)$ is measured.

\begin{figure}[ht]
\subfigure[]{
\includegraphics[width=0.23 \textwidth]{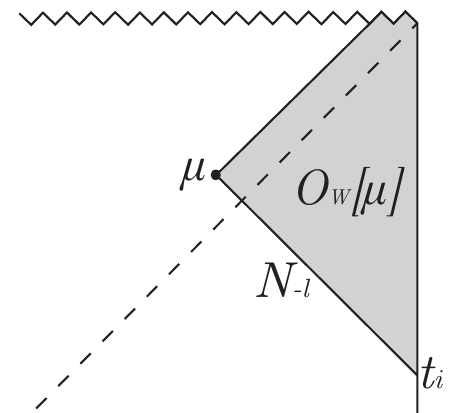}}
\subfigure[]{
\includegraphics[width=0.23 \textwidth]{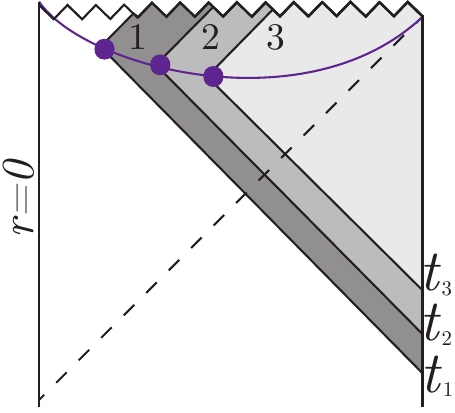}}

\caption{(a) We fire a null congruence $N_{-\ell}$ into the bulk from time $t=t_{i}$. The surface $\mu$ is the first cross-section of $N_{-\ell}$ with vanishing $k$ expansion.  We can recover all the data in $O_{W}[\mu]$, at least when the black hole is near equilibrium, by means of a ``simple experiment'' performed after time $t_{i}$. (b) A spacelike holographic screen (purple) has increasing area in a spacelike direction, going from 1 to 3. The corresponding outer wedges are nested, implying that the outer entropy must increase outwards. Similarly, the simple entropy must increase with $t$ from $t_{1}$ to $t_{3}$.}
\label{fig-examples}
\end{figure}

A simple experiment, by definition, can only access the subset of the bulk $F(t_{i})$ that is to the future of the boundary time $t_{i}$. When the spacetime has a black hole, turning on simple sources can shift the location of any event horizon $H$ in the spacetime~\cite{SheSta14}. However, the event horizon must always remain outside of any marginally trapped surface (assuming the Null Energy Condition)~\cite{HawEll,Wald}. Therefore, if $\mu$ is a marginally trapped surface on $N(t_{i})$, the boundary of $F(t_{i})$, a simple experiment can access at most the outer wedge $O_{W}[\mu]$. Note that by causality, turning on simple sources cannot modify the fact that $\mu$ is marginally trapped (a similar argument was given for extremal surfaces in~\cite{EngWal14}). See Fig.~\ref{fig-examples}(a).  It immediately follows that
\begin{equation} \label{eq:SimpleIneq}
S^{\mathrm{(simple)}}(t_{i})\ge S^{\mathrm{(outer)}}[\mu].
\end{equation}

If $N(t_{i})$ contains more than one marginally trapped surface, we restrict attention the earliest (i.e. outermost) one. This guarantees that $\mu$ is in fact an apparent horizon. We propose that in this case, the inequality~\eqref{eq:SimpleIneq} is saturated. In other words \textit{the simple entropy is the holographic dual of the area of the apparent horizon}.

We now show that this is true for a black hole that is approaching thermal equilibrium after time $t_{i}$. We may use the ``HKLL'' procedure~\cite{HamKab05,HamKab06, HeeMar, BalKraLaw98, BalKraLaw98b, BanDou98, Ben99} 
to reconstruct the `causal wedge'' $C_{W}[t_{i}]$ of $t_{i}$, i.e. the subset of $F(t_{i})$ outside of the event horizon~\cite{BouLei12, HubRan12}.  If no matter or gravitational radiation were falling across the event horizon $H$, it would be stationary; there would be no separation between $H$ and $\mu$, and we would be done. In order to reconstruct the data in $O_{W}[\mu]$, we must ensure that no matter falls across $H$ after $\mu$. 

Since $\mu$ is perturbatively close to the event horizon,~\cite{Kab11,HeeMar} allows us to map the matter fields falling across the event horizon to data on the boundary. We can therefore turn these fields ``off'' by adding suitable sources to the boundary after $t_{i}$. This has the effect of shifting the event horizon to the location of $\mu$, so that $C_{W}[t_{i}] = O_{W}[\mu]$.\footnote{When lightrays in $N(t_{i})$ intersect before reaching $\mu$, $O_{W}[\mu] \supset C_{W}[t_{i}]]$ since the past boundaries do not coincide.  However, $O_{W}[\mu]$ still lies in the domain of dependence of $C_{W}[t_{i}]]$ allowing reconstruction of the full data.~\cite{EngWalTA}.}
This shows that we can use HKLL to reconstruct the spacetime data arbitrarily close to $\mu$.  (Although to reconstruct points a distance $\epsilon$ from $\mu$, we need to wait a time of order $\ln(\epsilon^{-1})$ for the signal to reach the boundary.)  This shows that, order-by-order in small perturbation to a stationary black hole,
\begin{equation} \label{eq:SimpleIneq}
S^{\mathrm{(simple)}}(t_{i})= \frac{\mathrm{Area}[\mu]}{4G\hbar}.
\end{equation}
This is a new entry in the holographic dictionary, which we conjecture also holds for finite deviations from thermality.
\section{An Explanation for the Second Law}

A surface $\mathcal{H}$ foliated by marginally trapped surfaces and satisfying certain regularity conditions obeys an area law: the area of the marginally trapped surfaces increases with evolution along $\mathcal{H}$~\cite{Hay93, AshKri02,  GouJar06,BouEng15a,SanWei16}. In the case where the marginally trapped surfaces foliating $\mathcal{H}$ are apparent horizons, $\mathcal{H}$ must be spacelike~\cite{Hay93}, and are called trapping horizons~\cite{Hay93}, dynamical horizons~\cite{AshKri02, AshGal05}, or spacelike future holographic screens~\cite{BouEng15a}. The area law for these surfaces says that the area of slices of $\mathcal{H}$ increase going in an outward direction.

The spacelike holographic screen $\mathcal{H}$ is illustrated in Fig.~\ref{fig-examples}(b) in a collapsing black hole, where such objects are ubiquitous. The area increases in outwards evolution along apparent horizon slices of $\mathcal{H}$. The corresponding outer wedges are nested: evolving in the direction of increasing area corresponds to computing the outer entropy of progressively smaller outer wedges. This provides an immediate explanation for why the outer entropy increases along $\mathcal{H}$: evolution along $\mathcal{H}$ is the equivalent of maximizing the von Neumann entropy with progressively fewer constraints.

From a boundary perspective, the simple entropy increases for much the same reason, since as $t_i$ is increased, there are fewer simple experiments available.  It may seem odd that the simple entropy also allows measurements to be made at times \emph{after} $t_i$, but this is equivalent to saying that, for a coarse-graining scheme to have a Second Law, information cannot be discarded if it is going to become available later.  (Our very late time cutoff $t_f$, which is held constant as $t_i$ is increased, prevents us from having to worry about recurrences.)

\vskip .2cm
\noindent \textbf{Acknowledgments:} {\small It is a pleasure to thank R. Bousso, X. Dong, G. Horowitz, J. Maldacena, D. Marolf, F. Pretorius, J. Santos, D. Stanford, H. Verlinde, S. Weinberg, B. White, and E. Witten for helpful discussions.  The work of NE is supported in part by NSF grant PHY-1620059, while AW was supported by the Institute for Advanced Study, the Raymond and Beverly Sackler Foundation Fund, and 
NSF grant PHY-1314311.}

\section{Appendix: Constraint Equations}

Since we are imposing data on $N_{-k}$, we need to use the ``characteristic initial data formalism''~\cite{Ren90, BraDro95, ChoCru10, Luk12, Chr12, ChrPae12, ChrPae14}, which guarantees the existence of a solution\footnote{Luk~\cite{Luk12} only guarantees a local solution, but then presumably it is possible to deform the characterstic Cauchy slice into a nearby spacelike slice, guaranteeing existence and uniqueness of $M'$~\cite{Wald}.)
} if we satisfy the following constraint equations on $N_{-k}$ (one for each spacetime dimension $D$):
\begin{align}
&\nabla_k \theta_{k} = -\tfrac{1}{D-2}\theta_{k}^{2} - \varsigma_k^{2} - 8\pi G\,T_{kk}, \\
&\nabla_{k} \chi_{i}= -\theta_{k}\chi_{i} + \left(\tfrac{D-3}{D-2} \right)\nabla_{i} \theta_{k} - (\nabla\cdot \varsigma_{k})_{i} +8 \pi G T_{ik},\\    
&\nabla_k \theta_{\ell}= -\tfrac{1}{2} \mathcal{R} -2\nabla\cdot \chi -\theta_{\ell}\theta_{k} +2 \chi^{2} + 8\pi G\,T_{\ell k},
\end{align}
as well as the corresponding junction conditions which require $\theta_{k}$, $\chi_{i}$, and $\theta_{\ell}$ to be continuous.
Here $\varsigma_k$ is the shear tensor,
which is free data on $N_{-k}$;  $\mathcal{R}$ is the \emph{intrinsic} Ricci curvature of cross-sections of $N_{-k}$; $\chi_i$ is a $D-2$ component twist 1-form gauge field that tells you how much a normal vector gets boosted when transported in the transverse $i$-direction;  $T_{ab}$ is the stress tensor.  All quantities are defined on constant $v$-slices, where $v$ is an affine parameter defined on each null geodesic of $N_{-k}$, normalized so that $\nabla_k = \nabla_v$, and $k \cdot \ell = -1$.

We can solve these constraint equations for stationary $N_{-k}$ by stipulating that $\varsigma = \theta_k = T_{kk} = T_{ki} = 0$, while $\mathcal{R}$, $\chi_i$, $T_{\ell k}$ are constant along $v$.  The marginality condition $\theta_{k}[\mu] = 0$ ensures continuity of $\theta_{\ell}$ and $\theta_{k}$ on the junction between $N_{-k}$ and $O_{W}[\mu]$.  The shear is generically discontinuous across the junction, but that is not a problem for local evolution of the Einstein equation~\cite{LukRod12,LukRod13}. We assume without proof that evolution is possible with AdS boundary conditions. 

The above conditions on the stress tensor can be satisfied by reasonable matter fields.  For a minimally coupled scalar field $\phi$, take $\phi=$ constant in the $k$-direction; for a Maxwell field $A_{a}$, impose $\nabla_{k}A_{i}=0$ in the gauge $A_{k}=0$. In the Maxwell case there is one additional constraint equation for $\nabla_k F_{\ell k}$ that is satisfied if the current $j_{k}=0$.



Because $\mu$ is a apparent horizon, generically $\nabla_k \theta_{\ell}<0$ on $N_{-k}$ and $\theta_{\ell}[\mu]<0$.  It follows that there exists an extremal cross-section $X$ of $N_{-k}$ with $\theta_{\ell}=0$ (and $\theta_{k}=0$).  We can solve for the location of $X$:
\begin{equation}
0 = \theta_{\ell}[\mu] +\theta_{\ell,k}\,v +\Box\,v + 2\chi \cdot \nabla v,
\end{equation}
where $v$ is a function of the transverse directions.  
There is a unique solution to this equation, with $v < 0$ (see~\cite{AndMar05}).

To complete our spacetime $M'$, we invoked CPT-conjugation across the extremal surface $X$.  The junction conditions are satisfied at $X$ because $\theta_{\ell} = \theta_{k} = 0$ while $\chi_{i}$, $F_{\ell k}$, $A_{i}$ and $\phi$ are even under CPT; for more general matter fields, we expect that CPT-invariance ensures that this gluing is always possible.

\bibliographystyle{utcaps}
\bibliography{all}

\providecommand{\href}[2]{#2}\begingroup\raggedright\begin{thebibliography}{10}

\bibitem{Jay57a}
E.~T. Jaynes, ``Information Theory and Statistical Mechanics,''
  \href{http://dx.doi.org/10.1103/PhysRev.106.620}{{\em Phys. Rev.} {\bfseries
  106} (May, 1957) 620--630}.
  \url{https://link.aps.org/doi/10.1103/PhysRev.106.620}.

\bibitem{Jay57b}
E.~T. Jaynes, ``{Information Theory and Statistical Mechanics. II},''
\href{http://dx.doi.org/10.1103/PhysRev.108.171}{{\em Phys. Rev.} {\bfseries
  108} (1957) 171--190}.

\bibitem{GelHar06}
M.~Gell-Mann and J.~Hartle, ``{Quasiclassical Coarse Graining and Thermodynamic
  Entropy},'' \href{http://dx.doi.org/10.1103/PhysRevA.76.022104}{{\em Phys.
  Rev.} {\bfseries A76} (2007) 022104},
\href{http://arxiv.org/abs/quant-ph/0609190}{{\ttfamily arXiv:quant-ph/0609190
  [quant-ph]}}.

\bibitem{Bek72}
J.~D. Bekenstein, ``Black holes and the second law,''
{\em Nuovo Cim. Lett.} {\bfseries 4} (1972) 737--740.

\bibitem{Haw75}
S.~W. Hawking, ``Particle Creation By Black Holes,'' {\em Commun. Math. Phys.}
  {\bfseries 43} (1975) 199.

\bibitem{Haw71}
S.~W. Hawking, ``Gravitational radiation from colliding black holes,''
{\em Phys. Rev. Lett.} {\bfseries 26} (1971) 1344--1346.

\bibitem{BarCar73}
J.~M. Bardeen, B.~Carter, and S.~W. Hawking, ``The Four Laws of Black Hole
  Mechanics,'' {\em Commun. Math. Phys.} {\bfseries 31} (1973) 161.

\bibitem{Car08}
S.~Carlip, ``{Black Hole Thermodynamics and Statistical Mechanics},''
  \href{http://dx.doi.org/10.1007/978-3-540-88460-6_3}{{\em Lect. Notes Phys.}
  {\bfseries 769} (2009) 89--123},
\href{http://arxiv.org/abs/0807.4520}{{\ttfamily arXiv:0807.4520 [gr-qc]}}.

\bibitem{HawEll}
S.~W. Hawking and G.~F.~R. Ellis, {\em The large scale stucture of space-time}.
\newblock Cambridge University Press, Cambridge, England, 1973.

\bibitem{Hay93}
S.~Hayward, ``{General laws of black hole dynamics},''
\href{http://dx.doi.org/10.1103/PhysRevD.49.6467}{{\em Phys.Rev.} {\bfseries
  D49} (1994) 6467--6474}.

\bibitem{AshKri02}
A.~Ashtekar and B.~Krishnan, ``{Dynamical horizons: Energy, angular momentum,
  fluxes and balance laws},''
  \href{http://dx.doi.org/10.1103/PhysRevLett.89.261101}{{\em Phys.Rev.Lett.}
  {\bfseries 89} (2002) 261101},
\href{http://arxiv.org/abs/gr-qc/0207080}{{\ttfamily arXiv:gr-qc/0207080
  [gr-qc]}}.

\bibitem{GouJar06}
E.~Gourgoulhon and J.~L. Jaramillo, ``{A 3+1 perspective on null hypersurfaces
  and isolated horizons},''
  \href{http://dx.doi.org/10.1016/j.physrep.2005.10.005}{{\em Phys. Rept.}
  {\bfseries 423} (2006) 159--294},
\href{http://arxiv.org/abs/gr-qc/0503113}{{\ttfamily arXiv:gr-qc/0503113
  [gr-qc]}}.

\bibitem{BouEng15a}
R.~Bousso and N.~Engelhardt, ``{New Area Law in General Relativity},''
  \href{http://dx.doi.org/10.1103/PhysRevLett.115.081301}{{\em Phys. Rev.
  Lett.} {\bfseries 115} no.~8, (2015) 081301},
\href{http://arxiv.org/abs/1504.07627}{{\ttfamily arXiv:1504.07627 [hep-th]}}.

\bibitem{SanWei16}
F.~Sanches and S.~J. Weinberg, ``{Refinement of the Bousso-Engelhardt Area
  Law},'' \href{http://dx.doi.org/10.1103/PhysRevD.94.021502}{{\em Phys. Rev.}
  {\bfseries D94} no.~2, (2016) 021502},
\href{http://arxiv.org/abs/1604.04919}{{\ttfamily arXiv:1604.04919 [hep-th]}}.

\bibitem{RyuTak06}
S.~Ryu and T.~Takayanagi, ``{Holographic derivation of entanglement entropy
  from AdS/CFT},'' \href{http://dx.doi.org/10.1103/PhysRevLett.96.181602}{{\em
  Phys.Rev.Lett.} {\bfseries 96} (2006) 181602},
\href{http://arxiv.org/abs/hep-th/0603001}{{\ttfamily arXiv:hep-th/0603001
  [hep-th]}}.

\bibitem{HubRan07}
V.~E. Hubeny, M.~Rangamani, and T.~Takayanagi, ``{A Covariant holographic
  entanglement entropy proposal},''
  \href{http://dx.doi.org/10.1088/1126-6708/2007/07/062}{{\em JHEP} {\bfseries
  0707} (2007) 062},
\href{http://arxiv.org/abs/0705.0016}{{\ttfamily arXiv:0705.0016 [hep-th]}}.

\bibitem{LewMal13}
A.~Lewkowycz and J.~Maldacena, ``{Generalized gravitational entropy},''
  \href{http://dx.doi.org/10.1007/JHEP08(2013)090}{{\em JHEP} {\bfseries 1308}
  (2013) 090},
\href{http://arxiv.org/abs/1304.4926}{{\ttfamily arXiv:1304.4926 [hep-th]}}.

\bibitem{DonLew16}
X.~Dong, A.~Lewkowycz, and M.~Rangamani, ``{Deriving covariant holographic
  entanglement},'' \href{http://dx.doi.org/10.1007/JHEP11(2016)028}{{\em JHEP}
  {\bfseries 11} (2016) 028},
\href{http://arxiv.org/abs/1607.07506}{{\ttfamily arXiv:1607.07506 [hep-th]}}.

\bibitem{HeaTak}
M.~Headrick and T.~Takayanagi, ``{A Holographic proof of the strong
  subadditivity of entanglement entropy},''
  \href{http://dx.doi.org/10.1103/PhysRevD.76.106013}{{\em Phys.Rev.}
  {\bfseries D76} (2007) 106013},
\href{http://arxiv.org/abs/0704.3719}{{\ttfamily arXiv:0704.3719 [hep-th]}}.

\bibitem{Tho93}
G.~'t~Hooft, ``{Dimensional reduction in quantum gravity},'' in {\em {Salamfest
  1993:0284-296}}, pp.~0284--296.
\newblock 1993.
\newblock
\href{http://arxiv.org/abs/gr-qc/9310026}{{\ttfamily arXiv:gr-qc/9310026
  [gr-qc]}}.
\newblock

\bibitem{Sus95}
L.~Susskind, ``The World as a hologram,'' {\em J. Math. Phys.} {\bfseries 36}
  (1995) 6377--6396, \href{http://arxiv.org/abs/hep-th/9409089}{{\ttfamily
  hep-th/9409089}}.

\bibitem{EngWal17}
N.~Engelhardt and A.~C. Wall, ``{No Simple Dual to the Causal Holographic
  Information?},''
\href{http://arxiv.org/abs/1702.01748}{{\ttfamily arXiv:1702.01748 [hep-th]}}.

\bibitem{Sor97}
R.~D. Sorkin, ``{The statistical mechanics of black hole thermodynamics},''
\href{http://arxiv.org/abs/gr-qc/9705006}{{\ttfamily arXiv:gr-qc/9705006
  [gr-qc]}}.

\bibitem{BiaMye12}
E.~Bianchi and R.~C. Myers, ``{On the Architecture of Spacetime Geometry},''
  \href{http://dx.doi.org/10.1088/0264-9381/31/21/214002}{{\em Class. Quant.
  Grav.} {\bfseries 31} (2014) 214002},
\href{http://arxiv.org/abs/1212.5183}{{\ttfamily arXiv:1212.5183 [hep-th]}}.

\bibitem{FreMos13}
B.~Freivogel and B.~Mosk, ``{Properties of Causal Holographic Information},''
  \href{http://dx.doi.org/10.1007/JHEP09(2013)100}{{\em JHEP} {\bfseries 09}
  (2013) 100},
\href{http://arxiv.org/abs/1304.7229}{{\ttfamily arXiv:1304.7229 [hep-th]}}.

\bibitem{KelWal13}
W.~R. Kelly and A.~C. Wall, ``{Coarse-grained entropy and causal holographic
  information in AdS/CFT},''
  \href{http://dx.doi.org/10.1007/JHEP03(2014)118}{{\em JHEP} {\bfseries 1403}
  (2014) 118},
\href{http://arxiv.org/abs/1309.3610}{{\ttfamily arXiv:1309.3610 [hep-th]}}.

\bibitem{Bou99d}
R.~Bousso, ``The holographic principle for general backgrounds,'' {\em Class.
  Quant. Grav.} {\bfseries 17} (2000) 997,
\href{http://arxiv.org/abs/hep-th/9911002}{{\ttfamily hep-th/9911002}}.

\bibitem{Bou02}
R.~Bousso, ``The holographic principle,'' {\em Rev. Mod. Phys.} {\bfseries 74}
  (2002) 825,
\href{http://arXiv.org/abs/hep-th/0203101}{{\ttfamily hep-th/0203101}}.

\bibitem{SanWei16a}
F.~Sanches and S.~J. Weinberg, ``{Holographic entanglement entropy conjecture
  for general spacetimes},''
  \href{http://dx.doi.org/10.1103/PhysRevD.94.084034}{{\em Phys. Rev.}
  {\bfseries D94} no.~8, (2016) 084034},
\href{http://arxiv.org/abs/1603.05250}{{\ttfamily arXiv:1603.05250 [hep-th]}}.

\bibitem{NomSal16b}
Y.~Nomura, N.~Salzetta, F.~Sanches, and S.~J. Weinberg, ``{Toward a Holographic
  Theory for General Spacetimes},''
  \href{http://dx.doi.org/10.1103/PhysRevD.95.086002}{{\em Phys. Rev.}
  {\bfseries D95} no.~8, (2017) 086002},
\href{http://arxiv.org/abs/1611.02702}{{\ttfamily arXiv:1611.02702 [hep-th]}}.

\bibitem{Wal10QST}
A.~C. Wall, ``{The Generalized Second Law implies a Quantum Singularity
  Theorem},'' \href{http://dx.doi.org/10.1088/0264-9381/30/19/199501,
  10.1088/0264-9381/30/16/165003}{{\em Class.Quant.Grav.} {\bfseries 30} (2013)
  165003},
\href{http://arxiv.org/abs/1010.5513}{{\ttfamily arXiv:1010.5513 [gr-qc]}}.

\bibitem{BouEng15c}
R.~Bousso and N.~Engelhardt, ``{Generalized Second Law for Cosmology},''
  \href{http://dx.doi.org/10.1103/PhysRevD.93.024025}{{\em Phys. Rev.}
  {\bfseries D93} no.~2, (2016) 024025},
\href{http://arxiv.org/abs/1510.02099}{{\ttfamily arXiv:1510.02099 [hep-th]}}.

\bibitem{EngWal14}
N.~Engelhardt and A.~C. Wall, ``{Quantum Extremal Surfaces: Holographic
  Entanglement Entropy beyond the Classical Regime},''
  \href{http://dx.doi.org/10.1007/JHEP01(2015)073}{{\em JHEP} {\bfseries 01}
  (2015) 073},
\href{http://arxiv.org/abs/1408.3203}{{\ttfamily arXiv:1408.3203 [hep-th]}}.

\bibitem{FauLew13}
T.~Faulkner, A.~Lewkowycz, and J.~Maldacena, ``{Quantum corrections to
  holographic entanglement entropy},''
  \href{http://dx.doi.org/10.1007/JHEP11(2013)074}{{\em JHEP} {\bfseries 1311}
  (2013) 074},
\href{http://arxiv.org/abs/1307.2892}{{\ttfamily arXiv:1307.2892}}.

\bibitem{DonLew17}
X.~Dong and A.~Lewkowycz, ``{Entropy, extremality, euclidean variations, and
  the equations of motion},''
\href{http://arxiv.org/abs/1705.08453}{{\ttfamily arXiv:1705.08453 [hep-th]}}.

\bibitem{AMPS}
A.~Almheiri, D.~Marolf, J.~Polchinski, and J.~Sully, ``{Black Holes:
  Complementarity or Firewalls?},''
\href{http://arxiv.org/abs/1207.3123}{{\ttfamily arXiv:1207.3123 [hep-th]}}.

\bibitem{AMPSS}
A.~Almheiri, D.~Marolf, J.~Polchinski, D.~Stanford, and J.~Sully, ``{An
  Apologia for Firewalls},''
\href{http://arxiv.org/abs/1304.6483}{{\ttfamily arXiv:1304.6483 [hep-th]}}.

\bibitem{MatPlu11}
S.~D. Mathur and C.~J. Plumberg, ``{Correlations in Hawking radiation and the
  infall problem},'' \href{http://dx.doi.org/10.1007/JHEP09(2011)093}{{\em
  JHEP} {\bfseries 09} (2011) 093},
\href{http://arxiv.org/abs/1101.4899}{{\ttfamily arXiv:1101.4899 [hep-th]}}.

\bibitem{BraPir09}
S.~L. Braunstein, S.~Pirandola, and K.~Życzkowski, ``{Better Late than Never:
  Information Retrieval from Black Holes},''
  \href{http://dx.doi.org/10.1103/PhysRevLett.110.101301}{{\em Phys. Rev.
  Lett.} {\bfseries 110} no.~10, (2013) 101301},
\href{http://arxiv.org/abs/0907.1190}{{\ttfamily arXiv:0907.1190 [quant-ph]}}.

\bibitem{MarWhiTA}
D.~Marolf and B.~White, to appear.

\bibitem{Jac99}
T.~Jacobson, ``On the nature of black hole entropy,''
\href{http://arXiv.org/abs/gr-qc/9908031}{{\ttfamily gr-qc/9908031}}.

\bibitem{Sor05}
R.~D. Sorkin, ``{Ten theses on black hole entropy},''
  \href{http://dx.doi.org/10.1016/j.shpsb.2005.02.002}{{\em Stud. Hist. Phil.
  Sci.} {\bfseries B36} (2005) 291--301},
\href{http://arxiv.org/abs/hep-th/0504037}{{\ttfamily arXiv:hep-th/0504037
  [hep-th]}}.

\bibitem{JacMar05}
T.~Jacobson, D.~Marolf, and C.~Rovelli, ``{Black hole entropy: Inside or
  out?},'' \href{http://dx.doi.org/10.1007/s10773-005-8896-z}{{\em Int. J.
  Theor. Phys.} {\bfseries 44} (2005) 1807--1837},
\href{http://arxiv.org/abs/hep-th/0501103}{{\ttfamily arXiv:hep-th/0501103
  [hep-th]}}.

\bibitem{FreHub05}
B.~Freivogel, V.~E. Hubeny, A.~Maloney, R.~C. Myers, M.~Rangamani, and
  S.~Shenker, ``{Inflation in AdS/CFT},''
  \href{http://dx.doi.org/10.1088/1126-6708/2006/03/007}{{\em JHEP} {\bfseries
  03} (2006) 007},
\href{http://arxiv.org/abs/hep-th/0510046}{{\ttfamily arXiv:hep-th/0510046
  [hep-th]}}.

\bibitem{HsuRee08}
S.~D.~H. Hsu and D.~Reeb, ``{Unitarity and the {Hilbert} space of quantum
  gravity},'' \href{http://dx.doi.org/10.1088/0264-9381/25/23/235007}{{\em
  Class. Quant. Grav.} {\bfseries 25} (2008) 235007},
\href{http://arxiv.org/abs/0803.4212}{{\ttfamily arXiv:0803.4212 [hep-th]}}.

\bibitem{Mar08}
D.~Marolf, ``{Black Holes, AdS, and CFTs},''
  \href{http://dx.doi.org/10.1007/s10714-008-0749-7}{{\em Gen. Rel. Grav.}
  {\bfseries 41} (2009) 903--917},
\href{http://arxiv.org/abs/0810.4886}{{\ttfamily arXiv:0810.4886 [gr-qc]}}.

\bibitem{Wal12}
A.~C. Wall, ``{Maximin Surfaces, and the Strong Subadditivity of the Covariant
  Holographic Entanglement Entropy},''
  \href{http://dx.doi.org/10.1088/0264-9381/31/22/225007}{{\em
  Class.Quant.Grav.} {\bfseries 31} no.~22, (2014) 225007},
\href{http://arxiv.org/abs/1211.3494}{{\ttfamily arXiv:1211.3494 [hep-th]}}.

\bibitem{AndMet07}
L.~Andersson and J.~Metzger, ``{The Area of horizons and the trapped region},''
  \href{http://dx.doi.org/10.1007/s00220-008-0723-y}{{\em Commun. Math. Phys.}
  {\bfseries 290} (2009) 941--972},
\href{http://arxiv.org/abs/0708.4252}{{\ttfamily arXiv:0708.4252 [gr-qc]}}.

\bibitem{Mar14}
M.~Mars, ``{Stability of Marginally Outer Trapped Surfaces and Geometric
  Inequalities},''
\href{http://dx.doi.org/10.1007/978-3-319-06349-2_8}{{\em Fundam. Theor. Phys.}
  {\bfseries 177} (2014) 191--208}.

\bibitem{Pen65}
R.~Penrose, ``Gravitational Collapse and Space-Time Singularities,''
  \href{http://dx.doi.org/10.1103/PhysRevLett.14.57}{{\em Phys. Rev. Lett.}
  {\bfseries 14} (Jan, 1965) 57--59}.
  \url{https://link.aps.org/doi/10.1103/PhysRevLett.14.57}.

\bibitem{Wald}
R.~M. Wald, {\em General Relativity}.
\newblock The University of Chicago Press, Chicago, 1984.

\bibitem{Mal01}
J.~M. Maldacena, ``{Eternal black holes in anti-de Sitter},''
  \href{http://dx.doi.org/10.1088/1126-6708/2003/04/021}{{\em JHEP} {\bfseries
  04} (2003) 021},
\href{http://arxiv.org/abs/hep-th/0106112}{{\ttfamily arXiv:hep-th/0106112
  [hep-th]}}.

\bibitem{CzeKar12}
B.~Czech, J.~L. Karczmarek, F.~Nogueira, and M.~Van~Raamsdonk, ``{The Gravity
  Dual of a Density Matrix},''
  \href{http://dx.doi.org/10.1088/0264-9381/29/15/155009}{{\em
  Class.Quant.Grav.} {\bfseries 29} (2012) 155009},
\href{http://arxiv.org/abs/1204.1330}{{\ttfamily arXiv:1204.1330 [hep-th]}}.

\bibitem{HeaHub14}
M.~Headrick, V.~E. Hubeny, A.~Lawrence, and M.~Rangamani, ``{Causality \&
  holographic entanglement entropy},''
  \href{http://dx.doi.org/10.1007/JHEP12(2014)162}{{\em JHEP} {\bfseries 12}
  (2014) 162},
\href{http://arxiv.org/abs/1408.6300}{{\ttfamily arXiv:1408.6300 [hep-th]}}.

\bibitem{JafLew15}
D.~L. Jafferis, A.~Lewkowycz, J.~Maldacena, and S.~J. Suh, ``{Relative entropy
  equals bulk relative entropy},''
\href{http://arxiv.org/abs/1512.06431}{{\ttfamily arXiv:1512.06431 [hep-th]}}.

\bibitem{DonHar16}
X.~Dong, D.~Harlow, and A.~C. Wall, ``{Bulk Reconstruction in the Entanglement
  Wedge in AdS/CFT},''
\href{http://arxiv.org/abs/1601.05416}{{\ttfamily arXiv:1601.05416 [hep-th]}}.

\bibitem{FauLew17}
T.~Faulkner and A.~Lewkowycz, ``{Bulk locality from modular flow},''
\href{http://arxiv.org/abs/1704.05464}{{\ttfamily arXiv:1704.05464 [hep-th]}}.

\bibitem{PolSus99}
J.~Polchinski, L.~Susskind, and N.~Toumbas, ``{Negative energy, superluminosity
  and holography},'' \href{http://dx.doi.org/10.1103/PhysRevD.60.084006}{{\em
  Phys.Rev.} {\bfseries D60} (1999) 084006},
  \href{http://arxiv.org/abs/hep-th/9903228}{{\ttfamily arXiv:hep-th/9903228
  [hep-th]}}.
Expanded version replacing earlier hep-th 9902182.

\bibitem{Fre02}
B.~Freivogel, S.~B. Giddings, and M.~Lippert, ``{Toward a theory of
  precursors},'' \href{http://dx.doi.org/10.1103/PhysRevD.66.106002}{{\em
  Phys.Rev.} {\bfseries D66} (2002) 106002},
\href{http://arxiv.org/abs/hep-th/0207083}{{\ttfamily arXiv:hep-th/0207083
  [hep-th]}}.

\bibitem{SheSta14}
S.~H. Shenker and D.~Stanford, ``{Black holes and the butterfly effect},''
  \href{http://dx.doi.org/10.1007/JHEP03(2014)067}{{\em JHEP} {\bfseries 03}
  (2014) 067},
\href{http://arxiv.org/abs/1306.0622}{{\ttfamily arXiv:1306.0622 [hep-th]}}.

\bibitem{HamKab05}
A.~Hamilton, D.~N. Kabat, G.~Lifschytz, and D.~A. Lowe, ``{Local bulk operators
  in AdS/CFT: A Boundary view of horizons and locality},''
  \href{http://dx.doi.org/10.1103/PhysRevD.73.086003}{{\em Phys.Rev.}
  {\bfseries D73} (2006) 086003},
\href{http://arxiv.org/abs/hep-th/0506118}{{\ttfamily arXiv:hep-th/0506118
  [hep-th]}}.

\bibitem{HamKab06}
A.~Hamilton, D.~N. Kabat, G.~Lifschytz, and D.~A. Lowe, ``{Holographic
  representation of local bulk operators},''
  \href{http://dx.doi.org/10.1103/PhysRevD.74.066009}{{\em Phys.Rev.}
  {\bfseries D74} (2006) 066009},
\href{http://arxiv.org/abs/hep-th/0606141}{{\ttfamily arXiv:hep-th/0606141
  [hep-th]}}.

\bibitem{HeeMar}
I.~Heemskerk, D.~Marolf, J.~Polchinski, and J.~Sully, ``{Bulk and Transhorizon
  Measurements in AdS/CFT},''
  \href{http://dx.doi.org/10.1007/JHEP10(2012)165}{{\em JHEP} {\bfseries 10}
  (2012) 165},
\href{http://arxiv.org/abs/1201.3664}{{\ttfamily arXiv:1201.3664 [hep-th]}}.

\bibitem{BalKraLaw98}
V.~Balasubramanian, P.~Kraus, and A.~E. Lawrence, ``{Bulk versus boundary
  dynamics in anti-de Sitter space-time},''
  \href{http://dx.doi.org/10.1103/PhysRevD.59.046003}{{\em Phys. Rev.}
  {\bfseries D59} (1999) 046003},
\href{http://arxiv.org/abs/hep-th/9805171}{{\ttfamily arXiv:hep-th/9805171
  [hep-th]}}.

\bibitem{BalKraLaw98b}
V.~Balasubramanian, P.~Kraus, A.~E. Lawrence, and S.~P. Trivedi, ``{Holographic
  probes of anti-de Sitter space-times},''
  \href{http://dx.doi.org/10.1103/PhysRevD.59.104021}{{\em Phys. Rev.}
  {\bfseries D59} (1999) 104021},
\href{http://arxiv.org/abs/hep-th/9808017}{{\ttfamily arXiv:hep-th/9808017
  [hep-th]}}.

\bibitem{BanDou98}
T.~Banks, M.~R. Douglas, G.~T. Horowitz, and E.~J. Martinec, ``{AdS dynamics
  from conformal field theory},''
\href{http://arxiv.org/abs/hep-th/9808016}{{\ttfamily arXiv:hep-th/9808016
  [hep-th]}}.

\bibitem{Ben99}
I.~Bena, ``{On the construction of local fields in the bulk of AdS(5) and other
  spaces},'' \href{http://dx.doi.org/10.1103/PhysRevD.62.066007}{{\em Phys.
  Rev.} {\bfseries D62} (2000) 066007},
\href{http://arxiv.org/abs/hep-th/9905186}{{\ttfamily arXiv:hep-th/9905186
  [hep-th]}}.

\bibitem{BouLei12}
R.~Bousso, S.~Leichenauer, and V.~Rosenhaus, ``{Light-sheets and AdS/CFT},''
  \href{http://dx.doi.org/10.1103/PhysRevD.86.046009}{{\em Phys.Rev.}
  {\bfseries D86} (2012) 046009},
\href{http://arxiv.org/abs/1203.6619}{{\ttfamily arXiv:1203.6619 [hep-th]}}.

\bibitem{HubRan12}
V.~E. Hubeny and M.~Rangamani, ``{Causal Holographic Information},''
  \href{http://dx.doi.org/10.1007/JHEP06(2012)114}{{\em JHEP} {\bfseries 1206}
  (2012) 114},
\href{http://arxiv.org/abs/1204.1698}{{\ttfamily arXiv:1204.1698 [hep-th]}}.

\bibitem{Kab11}
D.~Kabat, G.~Lifschytz, and D.~A. Lowe, ``{Constructing local bulk observables
  in interacting AdS/CFT},''
  \href{http://dx.doi.org/10.1103/PhysRevD.83.106009}{{\em Phys.Rev.}
  {\bfseries D83} (2011) 106009},
\href{http://arxiv.org/abs/1102.2910}{{\ttfamily arXiv:1102.2910 [hep-th]}}.

\bibitem{EngWalTA}
N.~Engelhardt and A.~Wall, to appear.

\bibitem{AshGal05}
A.~Ashtekar and G.~J. Galloway, ``{Some uniqueness results for dynamical
  horizons},'' \href{http://dx.doi.org/10.4310/ATMP.2005.v9.n1.a1}{{\em
  Adv.Theor.Math.Phys.} {\bfseries 9} (2005) 1--30},
\href{http://arxiv.org/abs/gr-qc/0503109}{{\ttfamily arXiv:gr-qc/0503109
  [gr-qc]}}.

\bibitem{Ren90}
A.~D. Rendall, ``Reduction of the Characteristic Initial Value Problem to the
  Cauchy Problem and Its Applications to the {Einstein} Equations,''
  \href{http://dx.doi.org/10.1098/rspa.1990.0009}{{\em Proc. Roy. Soc. Lon. A}
  {\bfseries 427} no.~1872, (1990) 221--239}.

\bibitem{BraDro95}
P.~R. Brady, S.~Droz, W.~Israel, and S.~M. Morsink, ``{Covariant double null
  dynamics: (2+2) splitting of the Einstein equations},''
  \href{http://dx.doi.org/10.1088/0264-9381/13/8/015}{{\em Class. Quant. Grav.}
  {\bfseries 13} (1996) 2211--2230},
\href{http://arxiv.org/abs/gr-qc/9510040}{{\ttfamily arXiv:gr-qc/9510040
  [gr-qc]}}.

\bibitem{ChoCru10}
Y.~Choquet-Bruhat, P.~T. Chrusciel, and J.~M. Martin-Garcia, ``{The Cauchy
  problem on a characteristic cone for the {Einstein} equations in arbitrary
  dimensions},'' \href{http://dx.doi.org/10.1007/s00023-011-0076-5}{{\em
  Annales Henri Poincare} {\bfseries 12} (2011) 419--482},
\href{http://arxiv.org/abs/1006.4467}{{\ttfamily arXiv:1006.4467 [gr-qc]}}.

\bibitem{Luk12}
J.~Luk, ``{On the Local Existence for the Characteristic Initial Value Problem
  in General Relativity},''
\href{http://arxiv.org/abs/1107.0898}{{\ttfamily arXiv:1107.0898 [gr-qc]}}.

\bibitem{Chr12}
P.~T. Chrusciel, ``{The existence theorem for the general relativistic Cauchy
  problem on the light-cone},''
  \href{http://dx.doi.org/10.1017/fms.2013.8}{{\em SIGMA} {\bfseries 2} (2014)
  e10},
\href{http://arxiv.org/abs/1209.1971}{{\ttfamily arXiv:1209.1971 [gr-qc]}}.

\bibitem{ChrPae12}
P.~T. Chrusciel and T.-T. Paetz, ``{The Many ways of the characteristic Cauchy
  problem},'' \href{http://dx.doi.org/10.1088/0264-9381/29/14/145006}{{\em
  Class. Quant. Grav.} {\bfseries 29} (2012) 145006},
\href{http://arxiv.org/abs/1203.4534}{{\ttfamily arXiv:1203.4534 [gr-qc]}}.

\bibitem{ChrPae14}
P.~T. Chruściel and T.-T. Paetz, ``{Characteristic initial data and smoothness
  of Scri. I. Framework and results},''
  \href{http://dx.doi.org/10.1007/s00023-014-0364-y}{{\em Annales Henri
  Poincare} {\bfseries 16} no.~9, (2015) 2131--2162},
\href{http://arxiv.org/abs/1403.3558}{{\ttfamily arXiv:1403.3558 [gr-qc]}}.

\bibitem{LukRod12}
J.~Luk and I.~Rodnianski, ``{Local Propagation of Impulsive
  GravitationalWaves},'' \href{http://dx.doi.org/10.1002/cpa.21531}{{\em
  Commun. Pure Appl. Math.} {\bfseries 68} (2015) 511--624},
\href{http://arxiv.org/abs/1209.1130}{{\ttfamily arXiv:1209.1130 [gr-qc]}}.

\bibitem{LukRod13}
J.~Luk and I.~Rodnianski, ``{Nonlinear interaction of impulsive gravitational
  waves for the vacuum Einstein equations},''
\href{http://arxiv.org/abs/1301.1072}{{\ttfamily arXiv:1301.1072 [gr-qc]}}.

\bibitem{AndMar05}
L.~Andersson, M.~Mars, and W.~Simon, ``{Local existence of dynamical and
  trapping horizons},''
  \href{http://dx.doi.org/10.1103/PhysRevLett.95.111102}{{\em Phys. Rev. Lett.}
  {\bfseries 95} (2005) 111102},
\href{http://arxiv.org/abs/gr-qc/0506013}{{\ttfamily arXiv:gr-qc/0506013
  [gr-qc]}}.

\end{thebibliography}\endgroup
\end{document}